\documentclass[12pt]{article}
\usepackage{graphicx}
 
\begin{document}
\begin{center}
{\large {\bf Modern Physics Simulations} }\\[8ex]
{\sl J.  A.  L\'opez, M.  Suskavcevic and C. Velasco } \\
Department of Physics\\ 
University of Texas at El Paso, El Paso, Texas 79968-0515\\[7ex]
\end{center}

Key words: Modern Physics, Simulations, MBL.

PACS: 01.30Lb, 01.40R, 01.50H, 01.50J.

\begin{center} 
{\large\bf Abstract \\[3ex] }
\end{center} 

{\noindent
Modern physics is now a regular course for non-physics majors who do 
not have to take the accompanying laboratory.  This lack of an 
experimental component puts the engineering students at a disadvantage.  
A possible solution is the use of computer simulations to add a 
constructivist element to the class.  In this work we present a set of 
computer simulations of fundamental experiments, key to the teaching of 
modern physics, as well as their in-class implementation and 
assessment.  Preliminary results indicate that the use of these 
simulations produce a substantial increase of student comprehension.}

\newpage

\section {Introduction}

Most universities and colleges offer a regular course on modern 
physics to non-physics majors.  Since in the majority of these cases, 
a laboratory is not required for this class, this puts engineering 
students at a disadvantage for understanding key concepts in the 
foundation of today's technology.  Solving this problem is a 
complicated issue, as engineering degree plans are already loaded with 
many courses and labs, and the equipment needed for many modern 
physics experiments is expensive and difficult to maintain.

This led us to explore different alternatives.  Hands on activities 
have been developed for general physics courses, but not much for 
modern physics.  A big exception is the project CUPS, Consortium for 
Upper-level Physics Software, which developed a series of computer 
exercises covering most of the undergraduate physics curriculum, 
including modern physics$^{1}$.

Although the core of the CUPS project provides instructors and students 
with a tool to teach physics and to develop physical intuition, the 
exercises lack definite goals, and end up being used mainly as 
in-class demonstrations and not as part of a constructivist 
environment.  Some accompanying material has been developed to 
integrate CUPS simulations to student solo activities$^{2}$.

 Generally speaking, the purpose of an experimental laboratory is to 
train the student in relating input variables with output signals, 
thus identifying physical concepts.  If performed under a 
constructivist environment, the student also learns how to advance 
hypotheses, develop theories, etc.  all under a group format.  To have 
computer simulations helping with all these tasks, the simulations 
should mimic, as much as possible, a real experiment $^{3}$.
Some of the problems of making a simulation as real as possible have been 
discussed in recent publications$^{4}$.

Under this spirit, a project to design, develop, and assess computer 
versions of key modern physics experiments was initiated with support 
from the Division of Undergraduate Education of the National Science 
Foundation (Grant NSF DUE-9651026).  This work is a report of the 
first in-class use and assessment of some of the simulations developed.

\section {The simulations}

Designing computer simulations as a supplement of a modern physics 
class imposes a number of limitations.  First, to maintain the class 
structure, the simulations must be planned as one-hour activities to 
replace a lecture.  The topic of each of the simulations has to be 
strongly correlated with the material covered in class.  The 
simulations must mimic, as much as possible, the setup of the original 
experiment, including input and output variables.  And, to add a 
constructivist touch, enough degrees of freedom had to be added to the 
simulations to allow the students to {\rm find their own way} and to 
avoid a recipe-like format.

\begin{figure}[htb]
\begin{center}
\includegraphics[width=\linewidth]{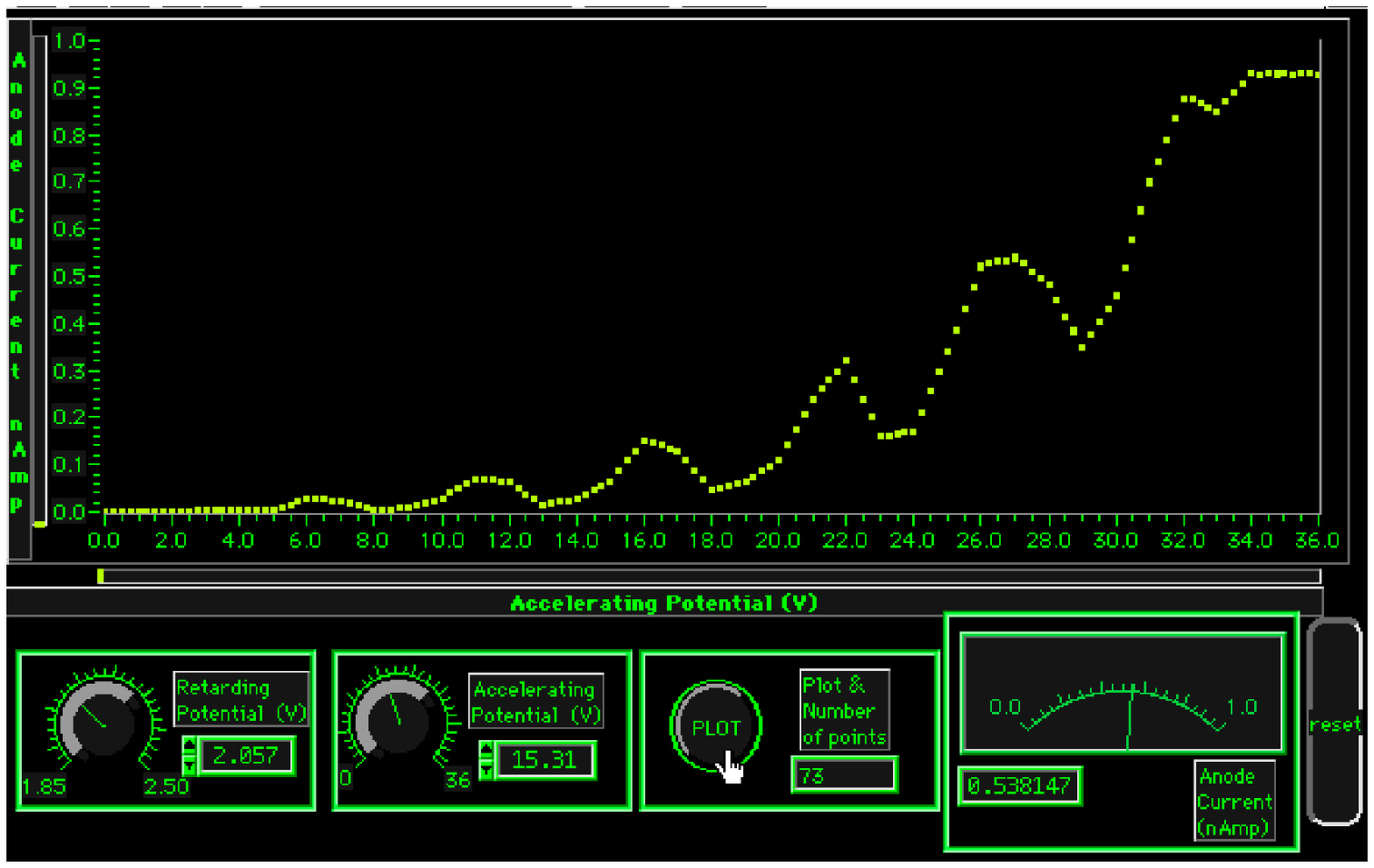}
\caption{Control panel for the simulation of the Frank-Hertz 
experiment.}
\label{fig1}
\end{center}
\end{figure}

The simulations developed included the classical experiments of 
Rutherford, Compton, Frank-Hertz, Davisson-Germer, Stern-Gerlack, 
Zeeman Effect, as well as the determination of Planck's constant.  All 
simulations have a one-page introduction to the experiment and a 
two-page activity guide.  A typical simulation consists of a full 
screen with controls and output devices.  As an illustration, fig.~\ref{fig1}
shows the control panel for the simulation of the Frank-Hertz 
experiment.

Along with the description of the experiment and activity guide, assessment 
instruments were developed for each of the simulations.  Pre- and post- 
examinations were designed to measure the impact of the simulation 
taken.

The first application of the simulations to modern physics students 
took place in the Fall of 1997 and Spring of 1998 at the University of 
Texas at El Paso (UTEP), and at the University of Wisconsin-Oshkosh 
(UWO).  In the following sections, as a case study, we describe one of 
the simulations in detail, and present assessment results of the use 
of several of these simulations.

\section{A study case: Davisson-Germer}

Davisson and Germer performed hundreds of experiments to demonstrate 
the wave-like properties of electrons scattering off a nickel crystal 
$^{5-7}$.  The electrons constituting an incident beam were 
thermally emitted from a tungsten ribbon and projected normally to a 
perfectly clean, gas-free, Ni target.  The intensity of scattering of 
this homogeneous beam of incident electrons was detected by movable 
detector in the range of colatitude angles from 0 to 90 degrees.  For 
a fixed voltage, maximum intensity peak appeared only at a certain 
colatitude angle.

Wavelengths associated with electrons using the De Broglie's relation 
$\lambda~=~h~/~m_{e}~v_{e}$ (with $h$ being Planck's constant, 
$m_{e}$ the electron mass and $v_{e}$ the experimentally set value for 
velocity) perfectly coincided, in a certain range of electrons' 
kinetic energies, with the wavelength values obtained by using Bragg's formula 
for maximum diffraction intensity $d \sin \theta = n \lambda$, where 
$\theta$ is the angle where maximum intensity is observed, $d$ is the 
interatomic spacing, and $n$ is the order of the diffraction maximum.  
Later on, experiments with different particles were performed and 
similar results were obtained.  The general conclusion was that particles 
behave like waves.

\subsection{Description of the simulation}

The simulation was designed to reproduce the experiment using a 
{\em black-box approach}.  Input and output variables, such as voltages, 
type of particles, etc.  were identified, and relationships among  
them were included as the {\em inner side} of the black-box.  

For the Davisson-Germer experiment, the input variables are 
the accelerating voltage, the type of particles (electrons), the 
diffracting crystal, and the detection angle.  Most of these variables 
can be changed by the user.  The only output variable is the scattering 
intensity which varies according to all input variables.  

In the computer program, the connection between the final intensity and 
the input variables comes from fitting original data, and not from a real- 
time simulation.  The data for the simulation was digitized from the 
original articles of Davisson and Germer $^{5-7}$ using experimentally obtained 
curves for seven different values of voltages from the range of 40 to 
68 Volts and for angles between 0 and 90 degrees.  Interpolation 
between any two curves makes possible to cover a larger range of 
voltages and colatitude angles.  In this way, the students have the 
same degrees of freedom that the original researchers had.  On their 
own, students must realize that to obtain meaningful results, an 
specific voltage must be fixed first, and that the detection angle 
must be varied many times.

\subsection{Assessment and data collection}

The assessment of the activity was designed to test the student's 
knowledge of the experiment before and after using the simulation.  
Again, taking the experiment as a black-box, elements of the experiment 
were identified to be included in the pre- and post-tests.  The assessment 
exams and the simulation were taken by the students after covering the 
material in class.  Two UTEP modern physics classes totalling over 50 
students participated in this study.  Due to the relatively small number of 
students, the possibility of using a control group was discarded.

\begin{figure}[htb]
\begin{center}
\includegraphics[width=\linewidth]{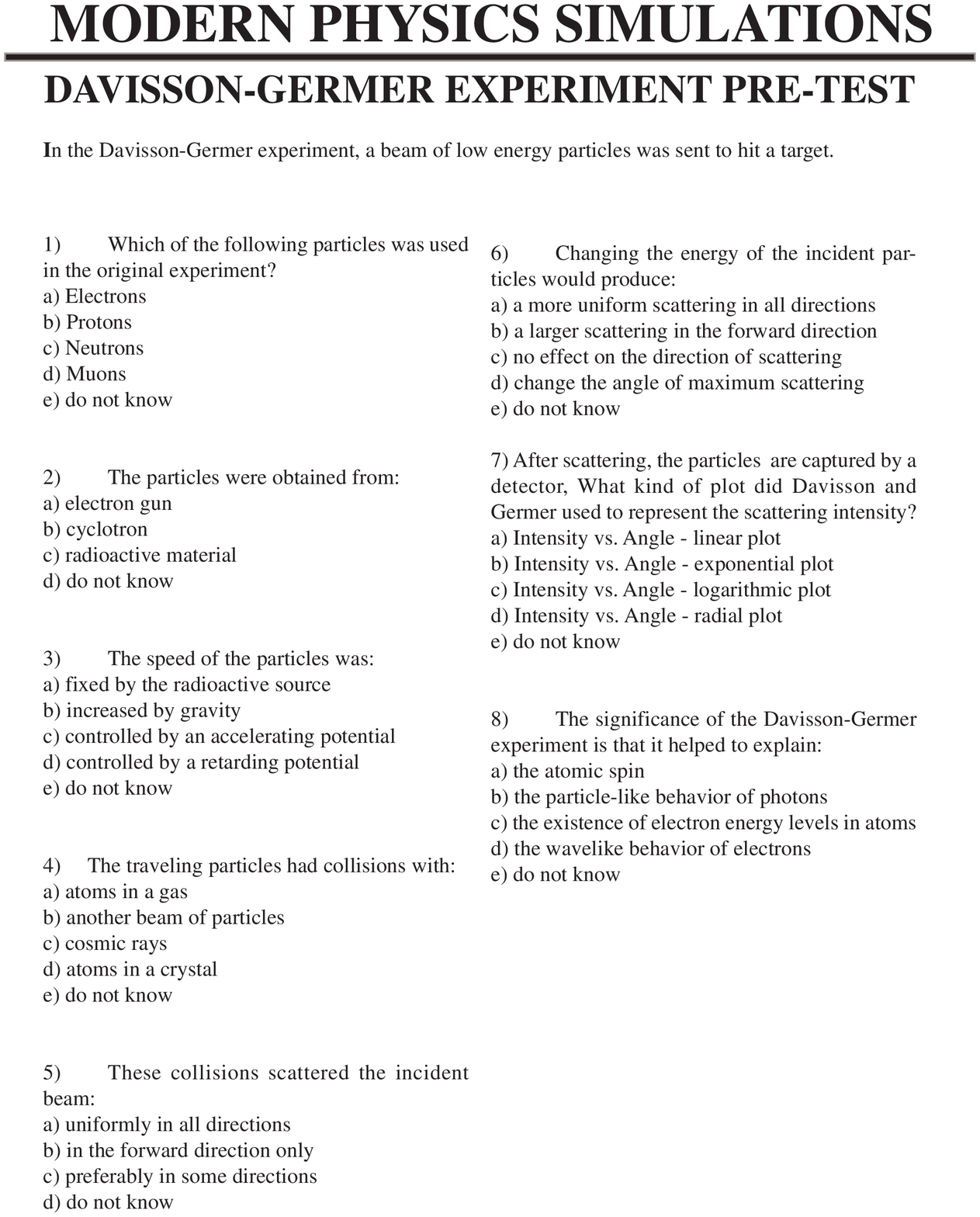}
\caption{Pre-exam for the simulation of the Davisson-Germer experiment.}
\label{fig2}
\end{center}
\end{figure}

\begin{figure}[htb]
\begin{center}
\includegraphics[width=\linewidth]{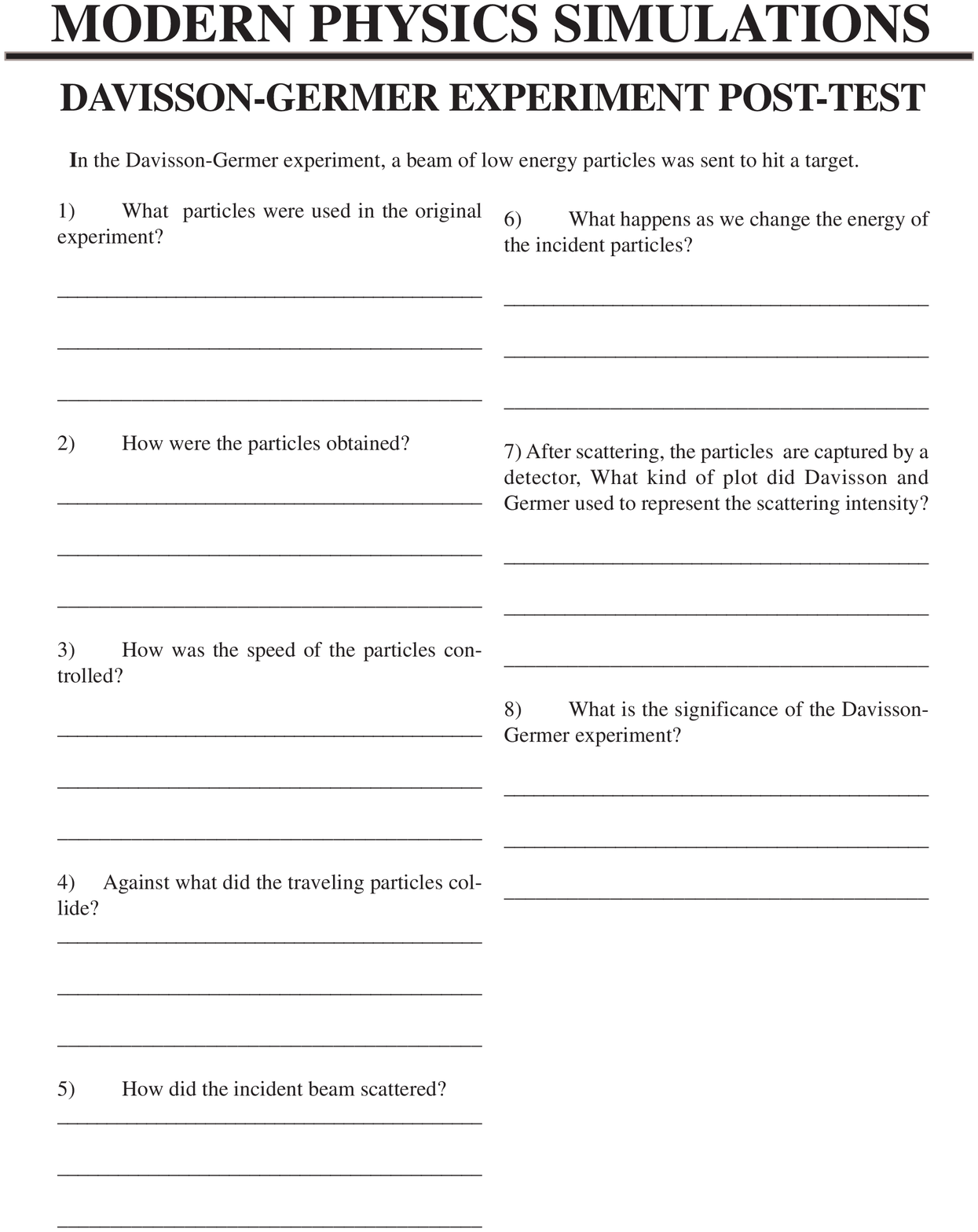}
\caption{Post-exam for the simulation of the Davisson-Germer experiment.}
\label{fig3}
\end{center}
\end{figure}

The pre- and post-exams for the Davisson-Germer simulation are shown in 
figs.~\ref{fig2} and~\ref{fig3}.  Most of the questions refer to basic knowledge of the 
physical ingredients of the experiment, and not to the meaning of 
the physical concepts.

The pre-test was administered in 10 minutes.  Afterwards 
the students were given the activity guide which consisted of a series 
of steps to illustrate the use of the simulation, and a series of 
open-ended questions to guide the students.  Working in groups of two, 
three or four, the students completed the simulation in about 45 
minutes, and immediately after running the simulation the students 
completed the post-test in about 10 more minutes.

As seen in figs.~\ref{fig2} and~\ref{fig3}, the pre- and post-examinations are much alike, 
except for the format.  This was to facilitate the correlation between 
exams on a question-to-question basis.  The grading of the essay-type 
post-exam took into account the depth of knowledge shown in the 
answers$^{8}$.

\subsection{Results}
\begin{figure}[htb]
\begin{center}
\includegraphics[width=\linewidth]{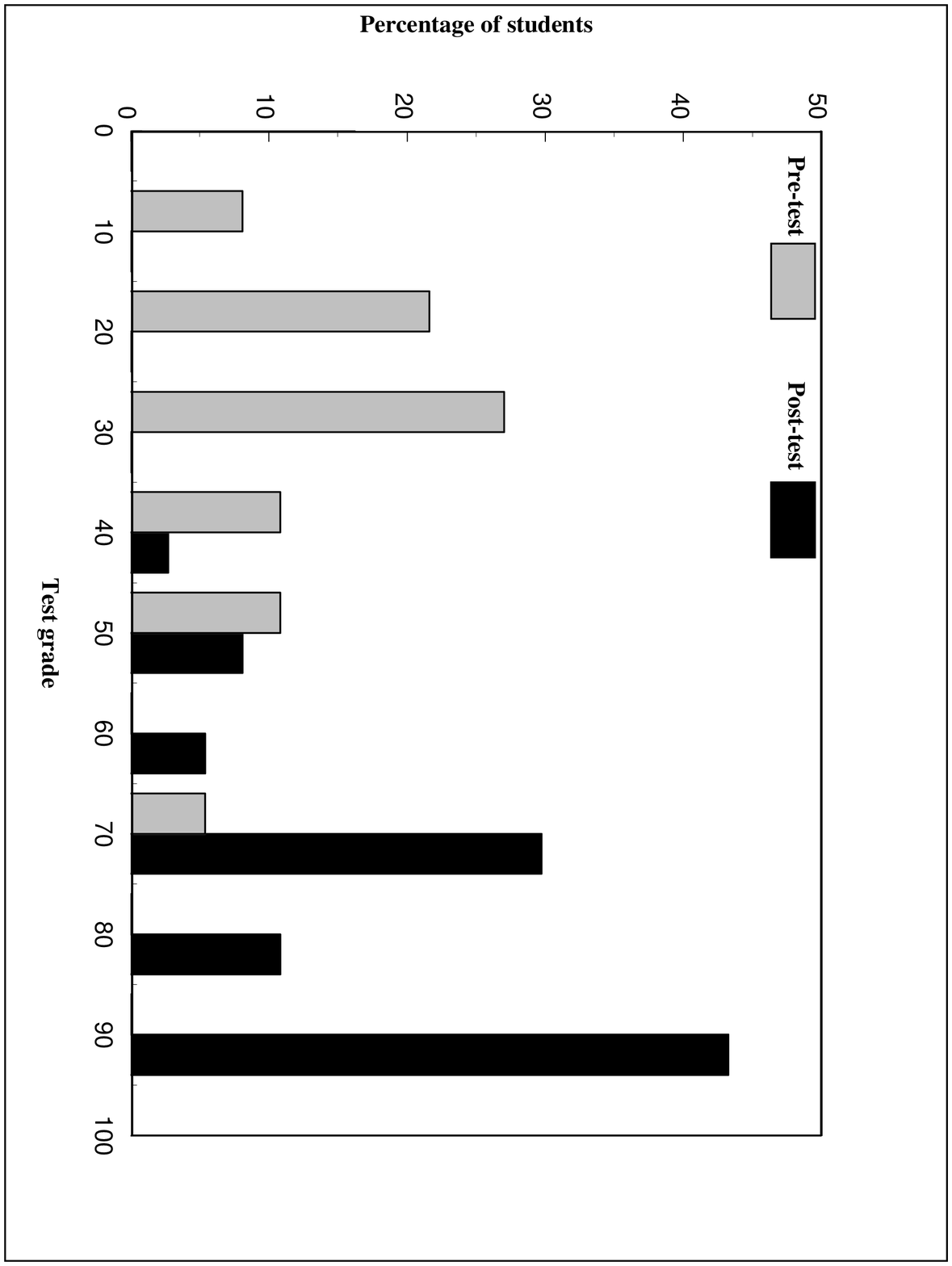}
\caption{Student performance in the pre- and post-exams.}
\label{fig4}
\end{center}
\end{figure}

\begin{figure}[htb]
\begin{center}
\includegraphics[width=\linewidth]{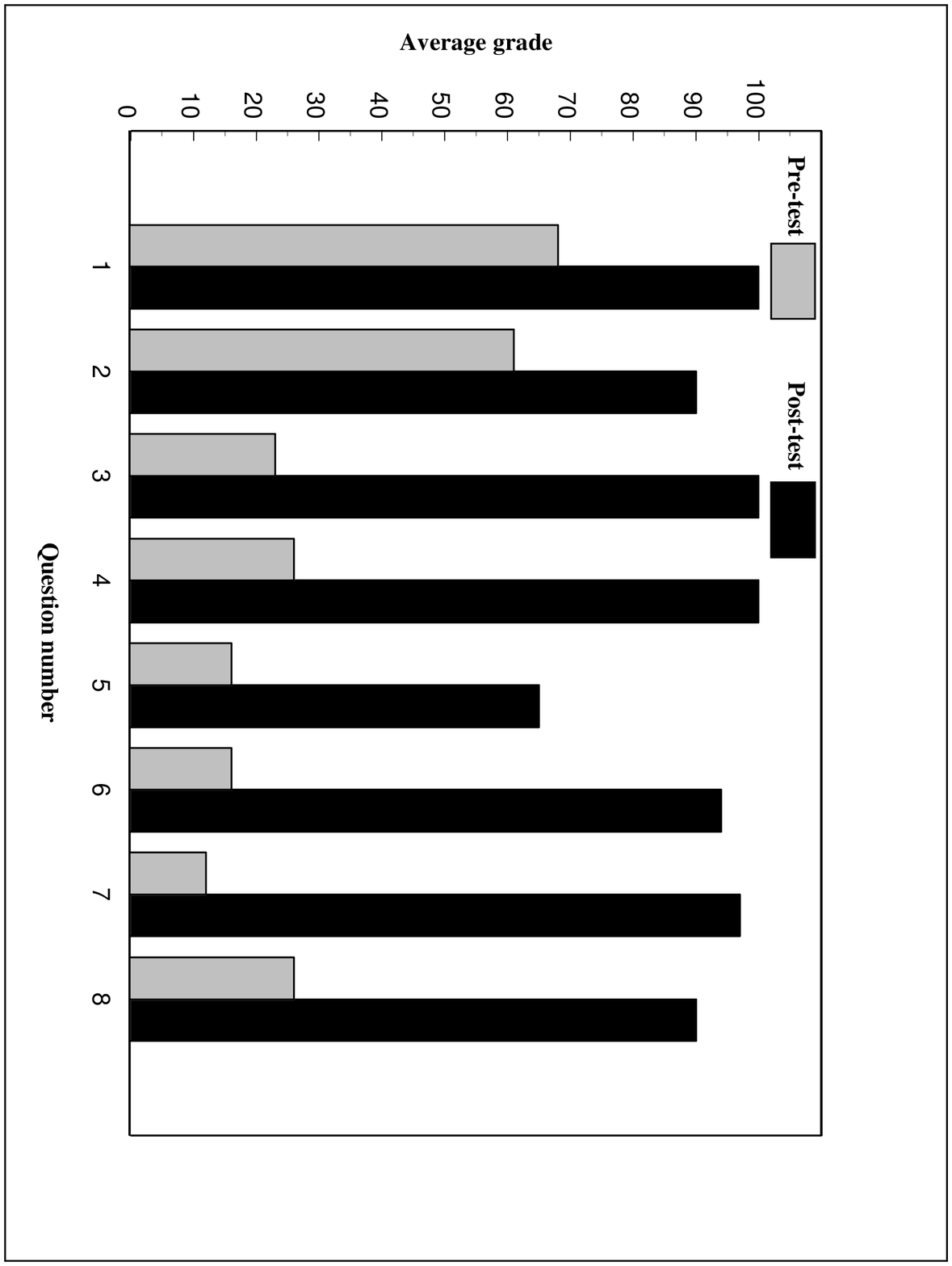}
\caption{Student performance per questions in the pre- and post-exams.}
\label{fig5}
\end{center}
\end{figure}

For the case of the Davisson-Germer experiment, the results 
regarding the student performance in the pre- and post-exams are displayed
in figs.~\ref{fig4}.  The same information, in terms of the questions, is shown 
in fig.~\ref{fig5}.

\section{Results from other simulations}
Altogether, four simulations were used and tested with the UTEP 
students, and one with one UWO modern physics class.  The simulations 
used at UTEP were the Rutherford, Compton, Frank-Hertz, and 
Davisson-Germer, and the one tested at UWO was the Frank-Hertz.

\begin{figure}[htb]
\begin{center}
\includegraphics[width=\linewidth]{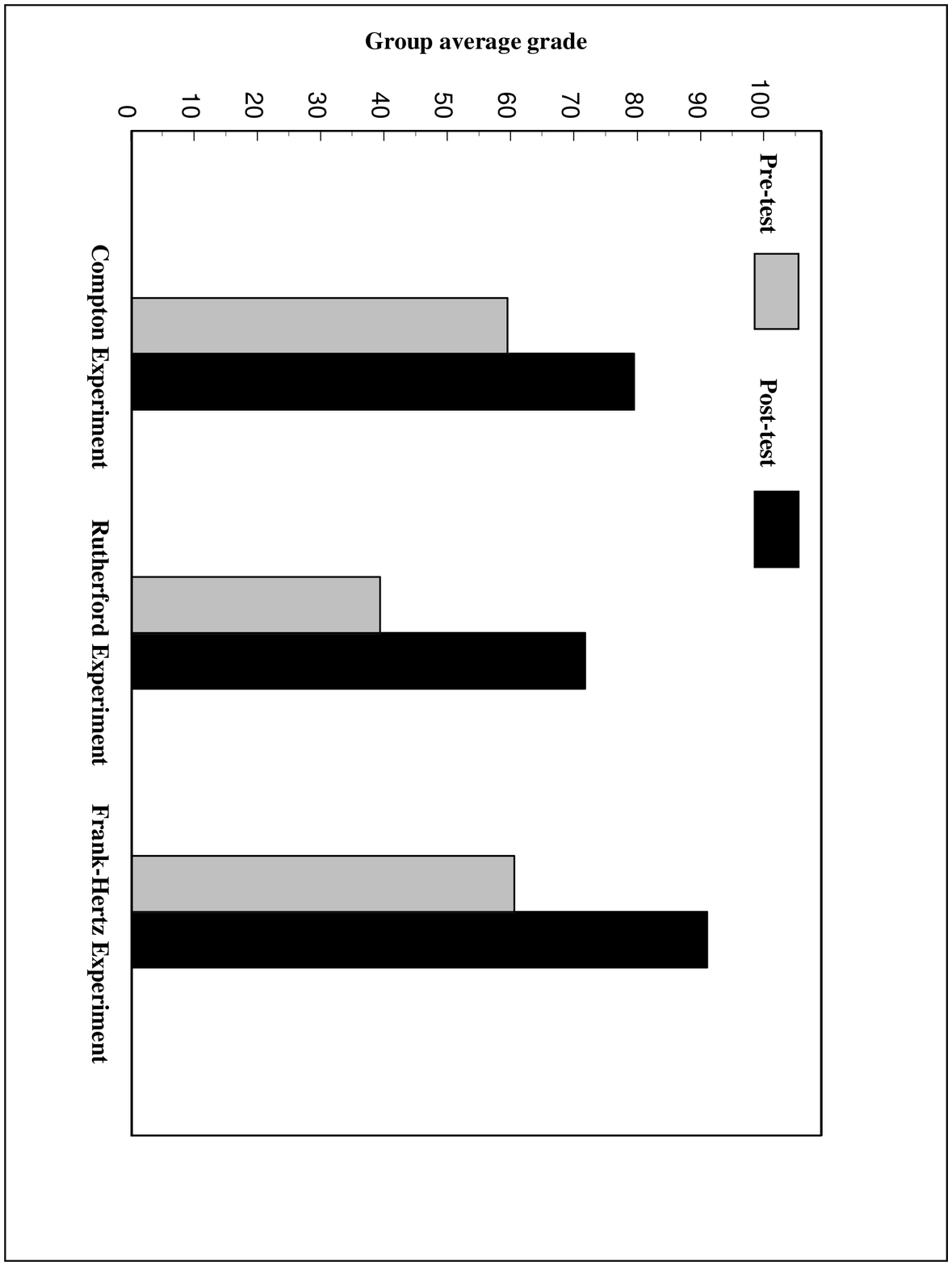}
\caption{Results for the pre- and post-exams performed 
for all the experiments.}
\label{fig6}
\end{center}
\end{figure}

Assessment analyses similar to the one described before were performed 
for all the experiments.  Results for the pre- and post-exams are 
presented in fig.~\ref{fig6}.  As before, the gain from the pre- to the post 
exams varied from 30\% to 50\%, much larger than the canonical $10$ to $20$\% 
generally obtained with traditional teaching methods, and on the higher end of
the percentages obtained with interactive engagement methods in introductory
physics courses$^{9}$. 

As an informal assessment, we collected anecdotal information from 
students, professors, and experts on the field.  Common student 
comments were:

\begin{itemize}
\item{\em ``The labs were useful and successful''}

\item{\em ``The laboratories offered in the physics department 
were very useful to me.  They gave me a more in-depth knowledge 
regarding the subject''}

\item {\em ``The class lecture gave me a good view behind the 
experiments, but the possibility of changing parameters and seeing the 
process of the experiment itself, gave me more details and helped me 
understand the material much better''}

\item {\em ``I feel more knowledgeable of the material for which I had 
the chance to run the experiments than those just presented in class.  
Also, it encouraged team discussion, and more ideas were brought up 
and shared among the group members''}.

\end{itemize}

\section{Conclusions and further directions}

To add an interactive element to modern physics courses, we designed, 
developed, and assessed computer versions of several key modern physics 
experiments.  Using the simulations as part of the class, students 
were tested before and after the use of the simulation.  We presented, 
as a study-case, the simulation, and the pre- and post-exams for the 
Davisson-Germer experiment.  Results for other simulations tested at 
UTEP and at UWO were also presented.

In conclusion, the use of the simulations appears to be very benefitial 
for the students.  Aside from extremely positive feedback from the 
participant instructors and students, the pre- and post-exams showed a 
significant increase in the understanding of the basic physics ingredients 
of the experiments.

\section{Acknowledgement}

This work was partially supported by the National Science Foundation 
under grant NSF DUE-9651026, and UTEP's NSF-MIE program.  We thank 
Dr.  R.  Fitzgerald, G.  Vandergrift, from UTEP, and Dr.  C.  Passow, 
from UWO, for allowing us to test the simulations in their modern 
physics classes.  We acknowledge interesting discussions with Dr.  R.  
Donangelo, from Universidad Federal do Rio de Janeiro, and Drs.  Pat 
Heller and Ken Heller, from the University of Minnesota.  We also 
thank students J.  Correa and M.  Cort\'es for helping 
with the programming and assessment of some of the simulations.

\newpage

\section*{List of figures}
\newcounter{papasconqueso}

\begin{list}
{Figure \arabic{papasconqueso}.}{\usecounter{papasconqueso} 
\setlength{\rightmargin}{\leftmargin}}

\item Control panel for the Frank-Hertz experiment.
\item Davisson-Germer pre-test.  
\item Davisson-Germer post-test.
\item Grade distributions for the pre- and post-tests of the Davisson-Germer simulation.
\item Average grade distribution for the pre- and post-tests per question of the 
Davisson-Germer simulation.
\item Group average grade distributions for the pre- and post-tests for other simulations.
\end{list}

\end{document}